\documentclass[12pt,twoside]{article}
\usepackage{fleqn,espcrc1}

\usepackage{graphicx}

\title{Strange Dibaryons in Neutron Stars and in Heavy-Ion Collisions}

\author{J\"urgen Schaffner-Bielich%
\address{RIKEN BNL Research Center, Physics Department, 
Brookhaven National Laboratoy, Upton, NY 11973, USA}%
\thanks{I thank RIKEN, BNL, and the U.S. Department of Energy for providing
  the facilities essential for the completion of this work.}} 
       
\begin{document}

\maketitle

\begin{abstract}
The formation of dibaryons with strangeness are discussed for the interior of
neutron stars and for central relativistic heavy-ion collisions. We derive
limits for the properties of H-dibaryons from pulsar data. Signals for the
formation of possible bound states with hyperons at BNL's Relativistic
Heavy-Ion Collider (RHIC) are investigated by studying their weak decay
patterns and production rates.  
\end{abstract}

\section{HYPERONS AND DIBARYONS IN NEUTRON STARS}

There has been a lot of speculations about the appearance of new particles in
the interior of neutron stars. A traditional neutron star consists of
neutrons, protons, and leptons only. Cameron suggested first in 1959, that hyperons will
also appear at high baryon density \cite{Cameron59}. A possible phase transition
to quark matter was conjectured shortly after the quark model was introduced
\cite{Ivan65}. Pion condensation \cite{Migdal78} as well as kaon condensation
were considered later \cite{KN86}. Finally, strange stars, built out of
absolutely stable strange quark matter, have been studied within the MIT bag
model \cite{Haensel86,Alcock86}. The appearance of one or the other exotic
phase in the core of neutron stars is still a matter of active debates. 
Nevertheless, recent developments in the field indicate,
that hyperons are probably the first exotic particle to appear in neutron star 
matter at twice normal nuclear density. This result was consistently derived within
various, different models: nonrelativistic potential models
\cite{Balberg97}, quark-meson coupling model \cite{Pal99}, relativistic
mean-field models \cite{Glen85,Knorren95b,SM96}, relativistic Hartree-Fock
\cite{Huber98}, Brueckner-Hartree-Fock models \cite{Baldo00,Vidana00}, and
chiral effective Lagrangians \cite{Hanauske00}.
Hence, the onset of hyperons at $2\rho_0$ seems to be rather insensitive to
the underlying details of the hyperon-nucleon interaction used.

If there are a lot of hyperons in the interior of neutron stars, then it might 
be also possible to form dibaryons with strangeness. Besides the famous
H-dibaryon proposed by Jaffe \cite{Jaffe77} built out of six quarks, the most
recent version of the Nijmegen potential predicts the existence of deeply
bound states of two hyperons, involving the $\Sigma$'s and $\Xi$'s with
maximum isospin \cite{Stoks99a}. Their results can be understood
in terms of the underlying SU(3) symmetry for the baryon-baryon interactions
(see \cite{Stoks99a}). 

\begin{figure}[thbp]
\centerline{\includegraphics[height=0.35\textheight]{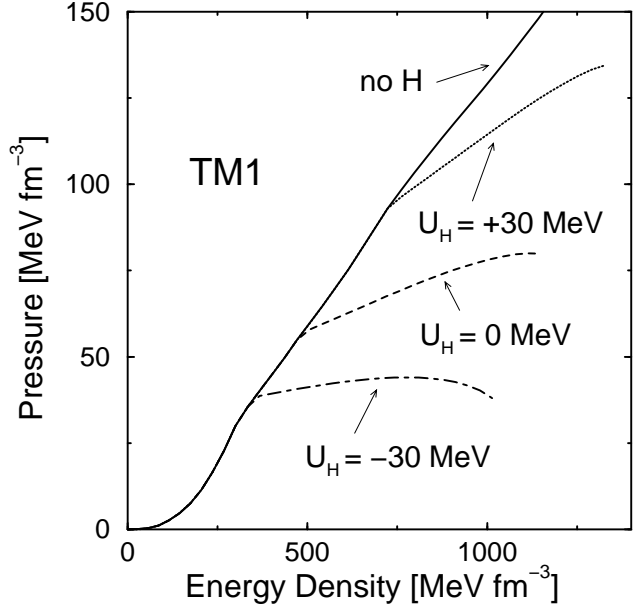}}
\vspace*{-1cm}
\caption{The equation of state of neutron star matter including a possible
  condensate of H-dibaryons. $U_H$ denotes the optical potential of the
  H-dibaryon at normal nuclear density. For an attractive potential, the
  equation of state is considerably softened (taken from \cite{GS98H}).}
\label{fig:eosH}
\end{figure}

Let us assume in the following, that there exist bound states with
hyperons. What would be the consequences for the properties of dense
matter, as present in the interior of neutron stars? Bound systems in dense
matter have been discussed before in connection with the liquid-gas phase
transition of nuclear matter. They dissolve at rather low density due to effects
in analogy to the Mott transition; deuterons disappear at $1/15\rho_0$ and $\alpha$
particles around $1/2 \rho_0$, as detailed e.g.\ in \cite{Roepke82}. A similar
transition will happen for bound states with hyperons. Hence, weakly bound
systems will have no influence on the equation of state of neutron star
matter --- unless, there is a deeply bound state which can be treated as a
quasiparticle. The H-dibaryon as well as heavier quark bags with strange
quarks (strangelets) are such deeply bound states. They can actually be formed as
precursor of the deconfinement phase transition in the core of neutron stars.
The general condition for the appearance of a strangelet in s-wave with baryon number $A$
and charge $Z$ is that the effective energy of the particle equals its chemical potential
\begin{equation}
E^*_s(k=0) = M_s + U(\rho) = A\times \mu_n - Z\times \mu_e
\quad .
\end{equation}
Here, $U(\rho)$ stands for the effective potential felt by the strangelet at
finite density. In order to estimate the effects of these exotic states, we focus on
the lightest possible one, the H-dibaryon with $A=2$ and $Z=0$. We use an
extended Relativistic Mean-Field Model which incorporates the H-particle in a
consistent scheme. The Lagrangian part for the H-dibaryon field reads
\begin{equation}
{\cal L}_H = D^*_\mu H^* D^\mu H - {m^*_H}^2 H^*H
\end{equation}
with a minimal coupling scheme for the scalar and vector meson fields
\begin{eqnarray}
D_\mu &=& \partial_\mu + i g_{\omega H} \cdot V_\mu \\
m^*_H &=& m_H + g_{\sigma H} \cdot \sigma
\quad .
\end{eqnarray}
For the coupling constant to the vector field, we choose the simple quark
counting rule as guidance, i.e.\ we set $g_{\omega H} = 4/3 g_{\omega N}$. We fix the 
scalar coupling to an optical potential at normal nuclear matter density in
the range
\begin{equation}
U_H (\rho_0) = g_{\sigma H} \cdot \sigma (\rho_0) + g_{\omega H} \cdot V_0
(\rho_0) = -30 \dots +30 \mbox{ MeV}
\quad .
\end{equation}
Figure \ref{fig:eosH} shows the resulting equation of state when using the
parameter set TM1 and including effects from hyperons (see \cite{GS98H} for
details). The equation of state turns out to be quite sensitive to the choice
of the optical potential of the H-dibaryon. The phase transition to the
H-dibaryon condensate is of second order in all cases studied. For an
attractive optical potential, $U_H(\rho_0)=-30$ MeV, the equation of state has 
a plateau-like behaviour, i.e.\ the pressure is very slowly rising with energy
density. 

\begin{figure}[thbp]
\centerline{\includegraphics[height=0.35\textheight]{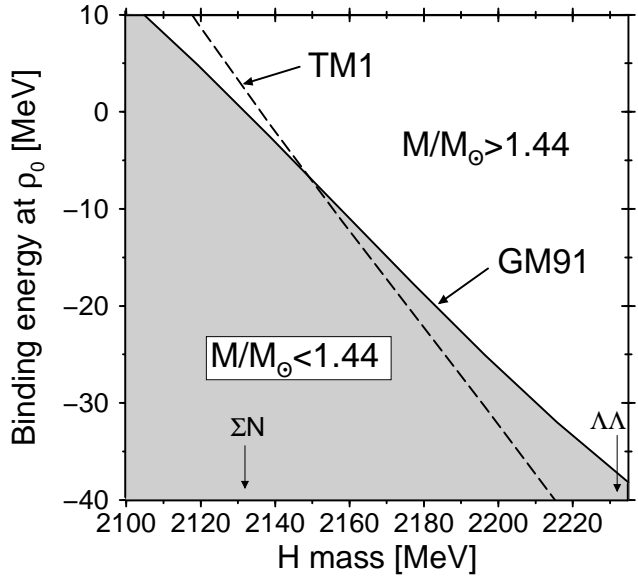}}
\vspace*{-1cm}
\caption{Neutron star constraints on the H-dibaryon from the mass of the
  Hulse-Taylor pulsar. A deeply bound H-dibaryon feeling an attractive
  potential in matter can be ruled out as the corresponding neutron stars have
  a too small maximum mass (from \cite{GS98H}).}
\label{fig:Hconstrain}
\end{figure}

The drastic softening of the equation of state has crucial
consequences for the maximum mass of a neutron star, which will be lowered
below the acceptable value of the Hulse-Taylor pulsar of $M=1.44M_\odot$. 
Hence, one can constrain the properties of the H-dibaryon in dense matter from 
pulsar data. These constraints are summarized in
fig.~\ref{fig:Hconstrain} as a function of the vacuum mass of the H-dibaryon
and its optical potential at $\rho_0$. The calculations have been done by
using two completely different parameterizations for the nucleon
and the hyperon coupling constants (denoted as set TM1 and GM91 in the figure)
and our findings seems to be rather insensitive to these choices.
According to these calculations, the shaded area in the lower left side
of the plot can be ruled out, as the corresponding neutron star equation of
state gives too low a maximum mass. Therefore, a deeply attractive potential
for the H-dibaryon with a mass close to the $\Lambda\Lambda$ threshold is
incompatible with the measured Hulse-Taylor pulsar mass. Also, a H-dibaryon 
mass below the $\Sigma$N threshold requires even a repulsive potential in matter
in our model calculation to get sufficiently heavy neutron stars. 

Finally, we remark that the existence of deeply bound states of two hyperons
can have an indirect effect on the properties of neutron stars, though. If the
hyperon-hyperon interaction is deeply attractive, it can give rise to a phase
transition to hyperonic matter \cite{SG00}. This phase transition can result 
in a drastically modified mass-radius relation for neutron stars, like
neutron star twins with similar masses but smaller radii than ordinary neutron 
stars \cite{SHSG00}. Hence, measurements of the mass and the radius of neutron 
stars will provide important insights into the equation of state 
and may reveal that there is indeed something strange going on at high density.

\section{STRANGE DIBARYONS IN RELATIVISTIC HEAVY-ION COLLISIONS}

There is the possibility to probe the hyperon-hyperon interaction in less
remote places than neutron stars. Terrestrial relativistic heavy-ion collisions will
produce dozens of hyperons in a single central collisions of two heavy
nuclei, and for RHIC even close to 200 hyperons are expected \cite{SMS00}.
The hyperons, sitting close in phase-space, can coalesce to form a
bound state or a resonance which is decaying afterwards. The final products
will be measured then in the detectors. 

The production rates for dibaryons
with hyperons at RHIC has been estimated in the coalescence model to be in the
range of $10^{-2}$ to $10^{-4}$ per single event. Effects from the details of
the dibaryon wavefunction have been found to be rather small \cite{SMS00}. 
The rapidity distribution is rather flat, so that dibaryons are also produced
at forward and backward rapidities where the decay length is considerably
longer than at midrapidity. The production rates can be enhanced dramatically
in two ways which are not taken into account in the above coalescence
estimates. If a quark-gluon plasma is formed, strange quark-antiquark pairs
are produced more abundantly so that the total initial number of produced strange
hadrons increases. If matter is created in a chirally restored phase, the
effective hyperon masses are reduced drastically which will result in an
enhanced hyperon-antihyperon yield. 

\begin{table}[tbp]
\caption{The hyperon weak decay amplitudes in SU(3)$_{\rm weak}$ compared to
  experimental data. All values are in units of $10^{-7}$.}   
\label{tab:weakamp}
\newcommand{\m}{\hphantom{$-$}}
\newcommand{\cc}[1]{\multicolumn{1}{r}{#1}}
\renewcommand{\tabcolsep}{2pc} 
\begin{tabular}{@{}lllll}
\hline
 & \multicolumn{2}{c}{parity violating} & \multicolumn{2}{c}{parity conserving} \cr
 & \cc{exp.} & \cc{SU(3)} & \cc{exp.} & \cc{SU(3)} \cr
\hline
$\Lambda\to p+\pi^-$ & \m3.25 & \m3.25 & \m22.1 & \m22.1 \cr
$\Lambda\to n+\pi^-$ & $-$2.37 & $-$2.30 & $-$16.0 & $-$15.6 \cr
$\Sigma^+\to n+\pi^+$ & \m0.13 & \m0.0 & \m42.2 & \m40.0 \cr
$\Sigma^+\to p+\pi^0$ & $-$3.27 & $-$3.33 & \m26.6 & \m28.3 \cr
$\Sigma^-\to n+\pi^-$ & \m4.27 & \m4.71 & $-$1.44 & \m0.0 \cr
$\Xi^0\to \Lambda+\pi^0$ & \m3.43 & \m3.19 & $-$12.3 & $-$11.7 \cr
$\Xi^-\to \Lambda+\pi^-$ & $-$4.51 & $-$4.51 & \m16.6 & \m16.6\\
\hline
\end{tabular}
\end{table}

To detect strange dibaryons, their weak decay patterns have to be
investigated. Starting point is the weak nonleptonic decay of hyperons in
vacuum.  Both, s-wave and p-wave amplitudes have been measured 
and correspond to a parity violating and a parity conserving amplitudes, respectively.
The standard approach for describing these amplitudes is by means of chiral
perturbation theory. The p-wave amplitudes are beyond leading order and
are usually derived within the pole model. The basic version fails to describe the
p-wave amplitudes. This constitutes the so called s-wave/p-wave puzzle for the 
weak nonleptonic decay of hyperons (for a
discussion and more elaborate ways to remedy the situation see
\cite{Holstein00} and references therein).  
We have found a simple way to parameterize both
amplitudes, s-wave and p-wave, in terms of pure SU(3) symmetry for the weak
interactions \cite{SMS00}. The effective Lagrangian involves the baryon octet
$B$, the pseudoscalar octet $P$, and the Gell-Mann matrix $\lambda_6$, which
ensures the $\Delta I=1/2$ rule and a change of hypercharge by one unit:
\begin{eqnarray}
{\cal L} &=& D \cdot {\rm Tr} \bar BB \left[P,\lambda_6\right] 
                    + F \cdot {\rm Tr} \bar B \left[P,\lambda_6\right] B
                    \nonumber \\ 
&&{} + G \cdot {\rm Tr} \bar B P \gamma_5 B \lambda_6
+ H \cdot {\rm Tr} \bar B \lambda_6 \gamma_5 B P 
+ J \cdot {\rm Tr} \bar B \{ P,\lambda_6\} \gamma_5 B 
\label{eq:Lagweak}
\quad .
\end{eqnarray}
The first two terms give the s-wave contributions, while the other three the
p-wave contributions. Table~\ref{tab:weakamp} compares the resulting amplitudes
with the measured ones when setting $D=4.72$, $F=-1.62$, $G=40.0$, $H=47.8$,
and $J=-7.1$ (in units of $10^{-10}$). 
We conclude that the above model as defined in eq.~(\ref{eq:Lagweak}) gives a reasonable
description of all amplitudes.  

\begin{figure}[t]
\centerline{\includegraphics[height=0.35\textheight]{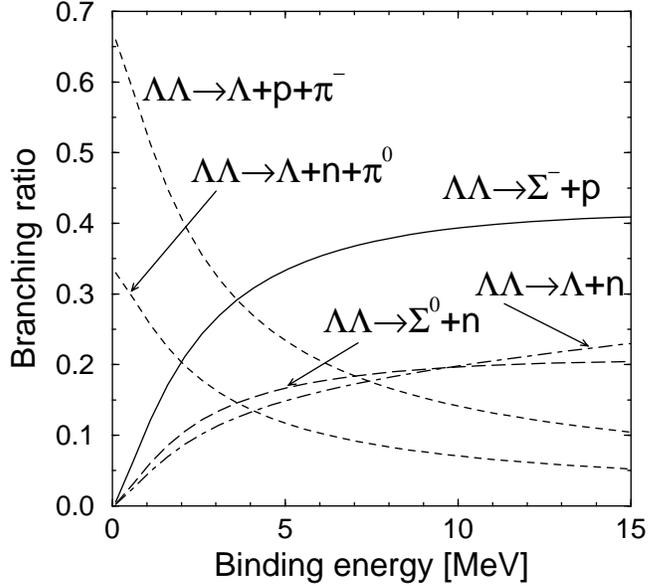}}
\vspace*{-1cm}
\caption{Branching ratios for a bound $\Lambda\Lambda$ state as a function of
  its binding energy. The nonmesonic decay mode $\Lambda\Lambda\to\Sigma^-+$p
  dominates for a binding energy of 4 MeV or more (taken from \cite{SMS00}).}
\label{fig:lala}
\end{figure}

The weak decays of possible dibaryons with hyperons are then calculated by
adopting a Hulthen-type wavefunction with varying binding energy. The
nonmesonic decay is described by pion- and kaon- exchange diagrams, where the
weak coupling constants are given by eq.~(\ref{eq:Lagweak}) and the strong
coupling constants by SU(6) symmetry. 
We note in passing, that the state-of-the-art calculations for the weak
nonmesonic decay of hypernuclei rely on the (wrong) pole model (see
\cite{Parreno97} and references therein). While this
will not affect the pion exchange diagrams, accidentally, it will certainly
alter the coupling constants for the kaon exchange diagrams. The use of the
SU(3) symmetric model might be interesting to pursue in this direction to
investigate possible effects, e.g.\ for the ratio of neutron- to
proton-induced decay which comes out too small in the standard approach.

\begin{figure}[t]
\centerline{\includegraphics[height=0.35\textheight]{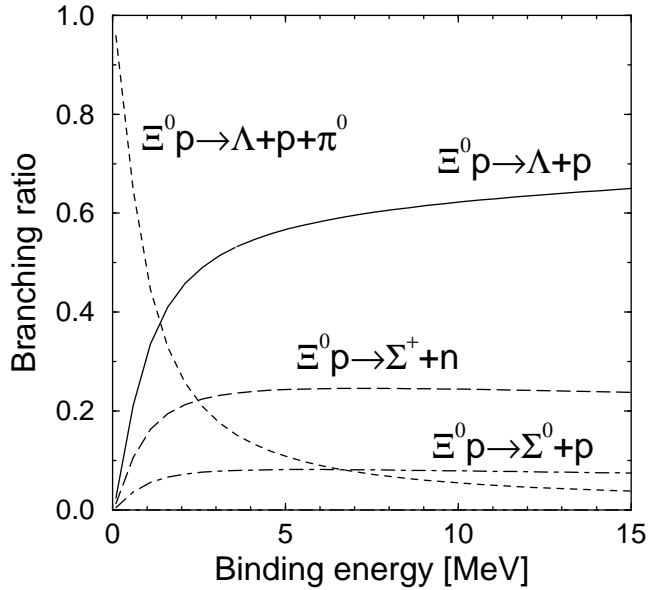}}
\vspace*{-1cm}
\caption{Branching ratios for a bound $\Xi^0$p state versus its binding
  energy. The decay mode $\Xi^0$p$\to\Lambda$p has the largest branching ratio
  for binding energies of 1.5 MeV or more (from \cite{SMS00}).}
\label{fig:xi0p}
\end{figure}

Figure~\ref{fig:lala} depicts the branching ratio for a bound $\Lambda\Lambda$ 
system versus the binding energy, and fig.~\ref{fig:xi0p} shows the case for a 
possible $\Xi^0$p dibaryon. The mesonic decay channel dominates for binding
energies of a few MeV. Then, for higher binding energies, the nonmesonic decay 
channels take over. For $\Lambda\Lambda$, the dominant nonmesonic decay mode is to a
$\Sigma^-$ and a proton, the same as for the hypothetical H-particle. 
The $\Xi^0$p decays mainly to a $\Lambda$ and a proton for binding energies of 
just 1.5 MeV and more. This weak decay has the same decay topology as the weak
decay of the $\Xi^-$ and the $\Omega^-$, and both particles have readily been 
measured in relativistic heavy-ion collisions \cite{Quercigh}. Other
interesting decay topologies occur for the doubly negatively charged dibaryons
$\Sigma^-\Sigma^-$, $\Sigma^-\Xi^-$, and $\Xi^-\Xi^-$, all of which have been predicted 
to be bound in the recent Nijmegen model \cite{Stoks99a}. 
Here, a negatively charged object decays into two negatively charged tracks --- a
unique decay prong which should be easily seen in tracking devices. The
calculated lifetimes of all dibaryons are lying just below the one for the free $\Lambda$ with
decay lengths in the range of $c\tau = 1$--5 cm. 

There are in principle three ways to detect possible short-lived strange
dibaryons (for an experimental investigation we refer to the 
detailed feasibility study of Coffin and Kuhn for the case of the H-dibaryon at the
STAR detector \cite{Coffin97}). 
Firstly, if the dibaryon is bound, one can look for exotic decays
in a TPC, like a charged track splitting into two charged ones. Secondly, if
the mass of the strange dibaryon is close to threshold, it will show up in the 
invariant mass spectrum of its decay products by using background
subtraction. Note, that this method is also 
sensitive to dibaryon resonances, if their decay widths are not too large
\cite{Paganis00}. Thirdly, the hyperon-hyperon interaction and possible broad
dibaryon resonances can be probed in 
correlation functions as the low momentum part is sensitive to final state
interactions \cite{Wang99,Ohnishi00}.

\end{document}